\documentclass{PoS}

\usepackage{amsmath}
\usepackage[utf8]{inputenc}
\DeclareMathOperator{\Tr}{Tr}

\title{Direct calculation of hadronic light-by-light scattering}

\ShortTitle{Direct calculation of hadronic light-by-light scattering}

\author{\speaker{Jeremy Green},
  Nils Asmussen,
  Oleksii Gryniuk,
  Georg von Hippel,
  Harvey~B.~Meyer,
  Andreas Nyffeler,
  and Vladimir Pascalutsa\\
  PRISMA Cluster of Excellence and Institut für Kernphysik,\\
  Johannes Gutenberg-Universität Mainz, Germany\\
  E-mail:
  \email{\{green,
    asmussen,
    gryniuk,
    hippel,
    meyerh,
    nyffeler,
    vladipas\}@kph.uni-mainz.de}
}

\abstract{We report calculations of hadronic light-by-light scattering
  amplitudes via lattice QCD evaluation of Euclidean four-point
  functions of vector currents. These initial results include only the
  fully quark-connected contribution. Particular attention is given to
  the case of forward scattering, which can be related via dispersion
  relations to the $\gamma^* \gamma^* \to \text{hadrons}$ cross
  section, and thus allows lattice data to be compared with
  phenomenology. We also present a strategy for computing the hadronic
  light-by-light contribution to the muon anomalous magnetic moment.}

\FullConference{The 33rd International Symposium on Lattice Field Theory\\
		 14--18 July  2015\\
		 Kobe International Conference Center, Kobe, Japan}

\begin{document}

\section{Introduction}

The anomalous magnetic moment of the muon,
$a_\mu\equiv(\frac{g-2}{2})_\mu$, is one of the most precise tests of
the Standard Model. Current values from experiment and theory are
\begin{equation}
  a_\mu = \begin{cases}
    116592089(63) \times 10^{-11} & \text{experiment~\cite{Agashe:2014kda,Bennett:2006fi}}\\
    116591790(65) \times 10^{-11} & \text{theory~\cite{Jegerlehner:2009ry}}
  \end{cases},
\end{equation}
which disagree by three standard deviations. Upcoming experiments at
Fermilab~\cite{Venanzoni:2014ixa} and J-PARC~\cite{Mibe:2011zz} plan
to reduce the experimental uncertainty by a factor of four or better,
which motivates efforts to similarly reduce the theoretical
uncertainty.

\begin{figure}
  \centering
  \includegraphics{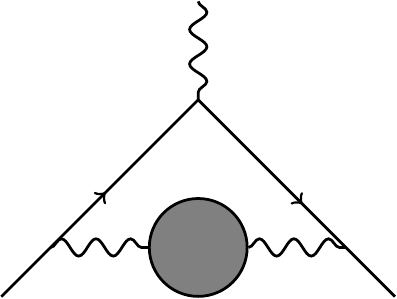}\hspace{5em}
  \includegraphics{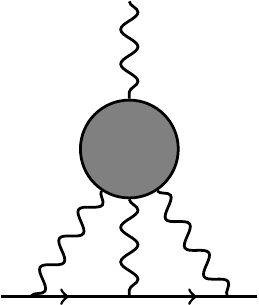}
  \caption{Leading-order contributions from the hadronic vacuum
    polarization (left) and hadronic light-by-light scattering (right)
    to $a_\mu$.}
  \label{fig:HVP_HLbL}
\end{figure}

The leading sources of theoretical uncertainty are hadronic
contributions: diagrams that include hadronic vacuum polarization and
hadronic light-by-light (HLbL) scattering
(Fig.~\ref{fig:HVP_HLbL}). The former can be determined using
$e^+e^-\to\text{ hadrons}$ cross sections measured in experiments,
whereas current predictions for the latter rely on models. Thus, a
reliable \emph{ab initio} calculation of the HLbL contribution using
lattice QCD would be useful for solidifying the Standard Model
prediction of $a_\mu$.  The first such calculations were done using
non-perturbative lattice QCD+QED by Blum \emph{et
  al.}~\cite{Blum:2014oka}. In this work we take a different route of
computing the HLbL amplitudes using non-perturbative lattice QCD; our
plan is to combine this with continuum perturbative QED to evaluate
the two-loop integrals and obtain the contribution to the muon $g-2$.

Our method provides an opportunity to study the HLbL scattering
amplitude by itself\footnote{Some of this work was previously reported
  in Ref.~\cite{Green:2015sra}.}. It could be used to compare
lattice calculations against phenomenology and provide an independent
check of the lattice methods. Conversely, it could be used for testing
models used for computing $a_\mu^\text{HLbL}$ by comparing them
against QCD calculations in a wide range of kinematics.

\section{Lattice evaluation of the four-point function}

With Wilson-type quarks, using the local and conserved vector
currents, and the contact operator,
\begin{equation}\label{eq:currents}
  \begin{aligned}
    J_\mu^l(x) &= Z_V \bar q(x) \gamma_\mu q(x) \\
    J_\mu^c(x) &= \frac{1}{2}\left[ 
      \bar q(x+\hat\mu) (\gamma_\mu+1) U_\mu^\dagger(x) q(x)
      + \bar q(x) (\gamma_\mu-1) U_\mu(x) q(x+\hat\mu) \right]\\
    T_\mu(x)  &= \frac{1}{2}\left[ 
      \bar q(x+\hat\mu) (\gamma_\mu+1) U_\mu^\dagger(x) q(x)
      - \bar q(x) (\gamma_\mu-1) U_\mu(x) q(x+\hat\mu) \right]
  \end{aligned}
\end{equation}
we compute the local-conserved-conserved-conserved four-point
function. In position space:
\begin{equation}
  \begin{aligned}
    \Pi^\text{pos}_{\mu_1\mu_2\mu_3\mu_4}(x_1,x_2,0,x_4) &= 
    \Bigl\langle J_{\mu_3}^l(0) \Bigl[ 
      J_{\mu_1}^c(x_1) J_{\mu_2}^c(x_2) J_{\mu_4}^c(x_4)
    + \delta_{\mu_1\mu_2} \delta_{x_1x_2} T_{\mu_1}(x_1) J_{\mu_4}^c(x_4) \\
    &\qquad 
    + \delta_{\mu_1\mu_4} \delta_{x_1x_4} T_{\mu_4}(x_4) J_{\mu_2}^c(x_2) 
    + \delta_{\mu_2\mu_4} \delta_{x_2x_4} T_{\mu_4}(x_4) J_{\mu_1}^c(x_1) \\
    &\qquad
    + \delta_{\mu_1\mu_4} \delta_{\mu_2\mu_4} \delta_{x_1x_4} \delta_{x_2x_4} J_{\mu_4}^c(x_4)
    \Bigr] \Bigr\rangle,
  \end{aligned}
\end{equation}
where the contact terms ensure that this satisfies the
conserved-current Ward identities using the backward lattice
derivative $\Delta$,
\begin{equation}
  \Delta_{\mu_1}^{x_1} \Pi^\text{pos}_{\mu_1\mu_2\mu_3\mu_4}
= \Delta_{\mu_2}^{x_2} \Pi^\text{pos}_{\mu_1\mu_2\mu_3\mu_4}
= \Delta_{\mu_4}^{x_4} \Pi^\text{pos}_{\mu_1\mu_2\mu_3\mu_4}
= 0.
\end{equation}
In our implementation, we have verified that these hold on each gauge
configuration.

\begin{figure}
  \centering
  \includegraphics[width=0.7\textwidth]{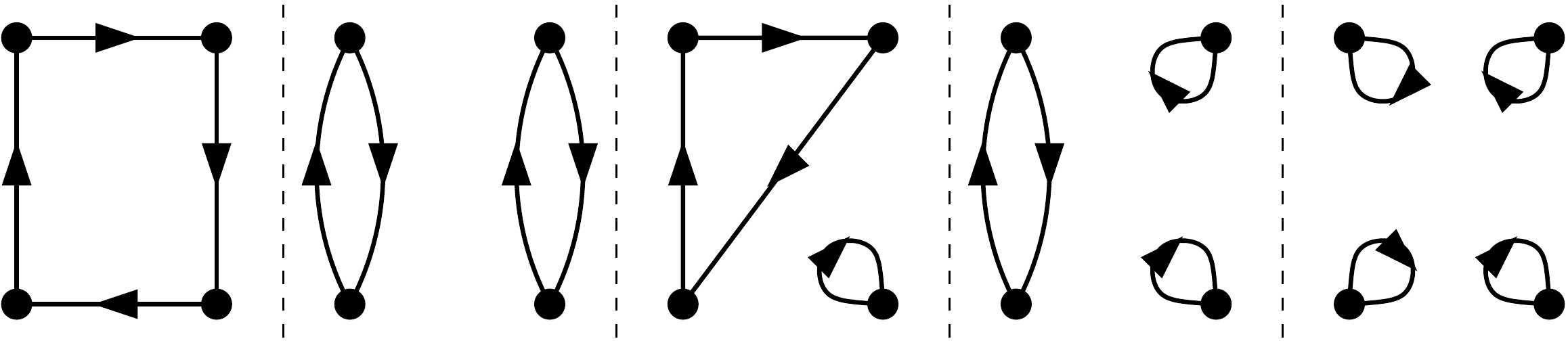}
  \caption{The five classes of quark contractions for four-point
    functions. In this work, we compute only the leftmost,
    fully-connected set of contractions.}
  \label{fig:contractions}
\end{figure}

\newcommand{\cV}{\mathcal{J}}
\newcommand{\cT}{\mathcal{T}}
There are five classes of quark contractions
(Fig.~\ref{fig:contractions}) required to evaluate the four-point
function. We evaluate only the fully-connected ones, with fixed
kernels summed over $x_1$ and $x_2$:
\begin{equation}
\Pi^{\text{pos}'}_{\mu_1\mu_2\mu_3\mu_4}(x_4;f_1,f_2)
= \sum_{x_1,x_2} f_1(x_2) f_2(x_2) \Pi^\text{pos}_{\mu_1\mu_2\mu_3\mu_4}(x_1,x_2,0,x_4).
\end{equation}
Using fixed kernels allows this to be evaluated using the combination
of a point-source propagator from the origin, and single- and
double-sequential propagators that contain one or both of the
kernels. If we define $\cV_\mu(x)$ and $\cT_\mu(x)$ to be the
point-split insertions in Eq.~(\ref{eq:currents}), then these three
kinds of propagators are
\begin{equation}
\begin{gathered}
  S_0(x) \equiv S(x,0), \quad
  S_{f\mu} \equiv S\left[\sum_x f(x)\cV_\mu(x) S_0 \right], \\
  S_{f\mu;g\nu} \equiv S\left[ \sum_x f(x)\cV_\mu(x) S_{g\nu}
    + \sum_x g(x)\cV_\nu(x) S_{f\mu}
    + \delta_{\mu\nu}\sum_x f(x)g(x)\cT_\mu(x) S_0 \right],
\end{gathered}
\end{equation}
and, noting that $\cV_\mu(x)$ is $\gamma_5$-antihermitian and
$\cT_\mu(x)$ is $\gamma_5$-hermitian, the connected four-point
function is obtained as
\begin{equation}
\begin{aligned}
  \Pi^{\text{pos}',\text{conn}}_{\mu_1\mu_2\mu_3\mu_4}(x_4;f_1,f_2) &=
  -\Bigl\langle\Tr\Bigl(\gamma_{\mu_3}\gamma_5\Bigl[
  S_{f_1^*\mu_1;f_2^*\mu_2}^\dagger\gamma_5\cV_{\mu_4}(x_4)S_0
  + S_0^\dagger\gamma_5\cV_{\mu_4}(x_4)S_{f_1\mu_1;f_2;\mu_2} \\
  &\qquad\qquad
  - S^\dagger_{f_2^*\mu_2}\gamma_5\cV_{\mu_4}(x_4)S_{f_1\mu_1}
  - S^\dagger_{f_1^*\mu_1}\gamma_5\cV_{\mu_4}(x_4)S_{f_2\mu_2} \\
  &\qquad\qquad
  + \delta_{\mu_1\mu_4}f_1(x_4)\bigl(
    S_0^\dagger\gamma_5\cT_{\mu_4}(x_4)S_{f_2\mu_2}
    - S_{f_2^*\mu_2}^\dagger\gamma_5\cT_{\mu_4}(x_4)S_0\bigr) \\
  &\qquad\qquad
  + \delta_{\mu_2\mu_4}f_2(x_4)\bigl(
    S_0^\dagger\gamma_5\cT_{\mu_4}(x_4)S_{f_1\mu_1}
    - S_{f_1^*\mu_1}^\dagger\gamma_5\cT_{\mu_4}(x_4)S_0\bigr) \\
  &\qquad\qquad
  + \delta_{\mu_1\mu_4}\delta_{\mu_2\mu_4}f_1(x_4)f_2(x_4)
  S_0^\dagger\gamma_5\cV_{\mu_4}(x_4)S_0
  \Bigr]\Bigr)\Bigr\rangle.
\end{aligned}
\end{equation}
Generically, this requires one point-source propagator $S_0$, sixteen
sequential propagators $S_{f\mu}$, $f\in\{f_1,f_2,f_1^*,f_2^*\}$, and
thirty-two double-sequential propagators $S_{f_1\mu;f_2\nu}$ and
$S_{f_1^*\mu;f_2^*\nu}$, for a total of 49 propagators, although this
number can be reduced in various special cases.

Finally, we obtain the momentum-space Euclidean four-point function
using plane waves:
\begin{equation}
  \Pi^E_{\mu_1\mu_2\mu_3\mu_4}(p_4;p_1,p_2)= \sum_{x_4} e^{-ip_4\cdot (x_4+\frac{a}{2}\hat\mu_4)}
 \Pi^{\text{pos}'}_{\mu_1\mu_2\mu_3\mu_4}(x_4;f_1,f_2)\Bigr|_{f_j(x)=e^{-ip_j\cdot (x+\frac{a}{2}\hat\mu_j)}}.
\end{equation}
Thus, by fixing $p_1$ and $p_2$ we can efficiently compute this for
all $p_4$.

\section{Lattice results}

We use three ensembles with $N_f=2$, $O(a)$-improved Wilson fermions
from CLS~\cite{Fritzsch:2012wq}. These have $a=0.063$~fm and pion
masses between 277 and 451~MeV. We keep only $u$ and $d$ quarks in the
electromagnetic current, so that in the above we actually use
$J_\mu^l=\frac{2}{3}\bar u\gamma_\mu u - \frac{1}{3} \bar d\gamma_\mu
d$,
etc, which amounts to applying an overall factor of $\frac{17}{81}$ to
the fully-connected four-point function for a single light quark. For
the results shown here, relatively few statistics were needed to
obtain a reasonable signal, and we used a maximum of 300 samples on
each ensemble.

The four-point function is a rather complicated object, and we focus
on the simplified case of forward kinematics ($p_1=-p_2$) where there
are only three kinematic invariants. We take the amplitude for
forward scattering of transversely polarized virtual photons,
\begin{equation}
  \mathcal{M}_{TT}(-Q_1^2,-Q_2^2,\nu) = \frac{e^4}{4} R_{\mu_1\mu_2}R_{\mu_3\mu_4}
  \Pi^E_{\mu_1\mu_2\mu_3\mu_4}(-Q_2;-Q_1,Q_1),
\end{equation}
where $\nu\equiv-Q_1\cdot Q_2$ and
\begin{equation}
  R_{\mu\nu} \equiv \delta_{\mu\nu} - \frac{1}{(Q_1\cdot Q_2)^2-Q_1^2Q_2^2} \left[
    (Q_1\cdot Q_2)(Q_{1\mu}Q_{2\nu} + Q_{1\nu}Q_{2\mu})
    - Q_1^2Q_{2\mu}Q_{2\nu} - Q_2^2Q_{1\mu}Q_{1\nu} \right]
\end{equation}
is a projector\footnote{We actually use the lattice momenta, $\hat
  Q_{i\mu} \equiv \frac{2}{a}\sin\frac{a}{2}Q_{i\mu}$.} onto the plane
orthogonal to $Q_1$ and $Q_2$.  At fixed spacelike $Q_1^2$ and
$Q_2^2$, this is related via a subtracted dispersion relation to the
helicity 0 and 2 cross sections for
$\gamma^*\gamma^*\to\text{hadrons}$, $\sigma_0$ and
$\sigma_2$~\cite{Pascalutsa:2012pr}:
\begin{equation}\label{eq:dispersion}
  \mathcal{M}_{TT}(-Q_1^2,-Q_2^2,\nu)-\mathcal{M}_{TT}(-Q_1^2,-Q_2^2,0)
= \frac{2\nu^2}{\pi} \int_{\nu_0}^\infty d\nu' 
\frac{\sqrt{\nu'^2-Q_1^2Q_2^2}}{\nu'(\nu'^2-\nu^2-i\epsilon)}
\left[\sigma_0(\nu') + \sigma_2(\nu')\right],
\end{equation}
where $\nu_0$ denotes the hadron-production threshold.
By itself, this relation is model-independent. As inputs from
experiment improve and the lattice calculations become more physical,
this will allow for systematically improved comparisons between
lattice and experiment.
For now, we use a phenomenological
model~\cite{Pascalutsa:2012pr,Green:2015sra} for $\sigma_0+\sigma_2$,
based on single-meson and non-resonant $\pi^+\pi^-$ final states.

\begin{figure}
  \centering
  \begin{minipage}{0.485\textwidth}
  \centering
  \includegraphics[width=\textwidth]{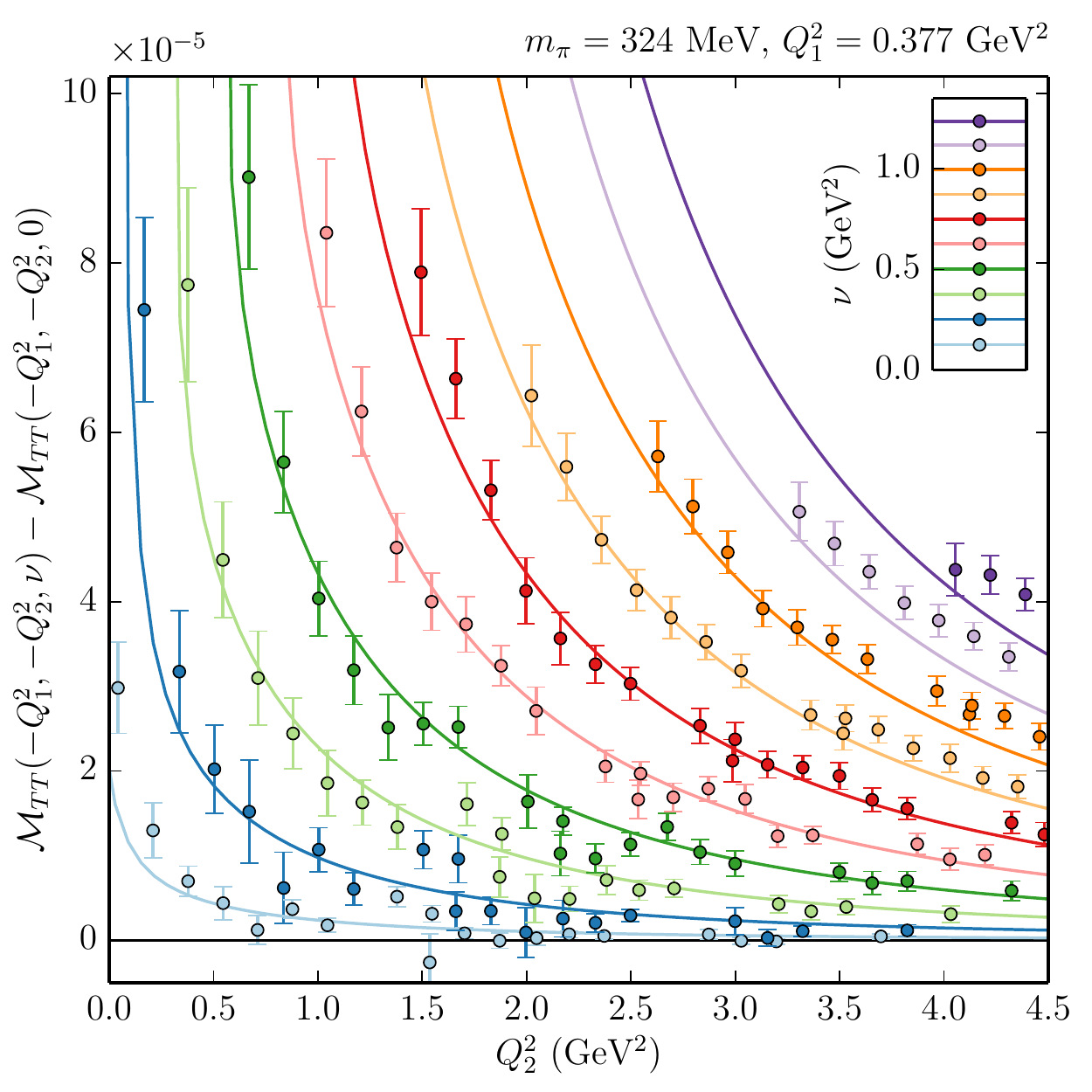}
  \caption{Forward scattering amplitude $\mathcal{M}_{TT}$ versus
    $Q_2^2$, for fixed $Q_1^2$ and pion mass, and a range of
    $\nu$. The points are lattice data and the curves result from
    applying the dispersion relation
    [Eq.~(\protect\ref{eq:dispersion})] to the model for the
    $\gamma^*\gamma^*\to\text{hadrons}$ cross section.}
  \label{fig:amp_fix_Q1_mpi}
  \end{minipage}\hspace{0.02\textwidth}
  \begin{minipage}{0.485\textwidth}
  \centering
  \includegraphics[width=\textwidth]{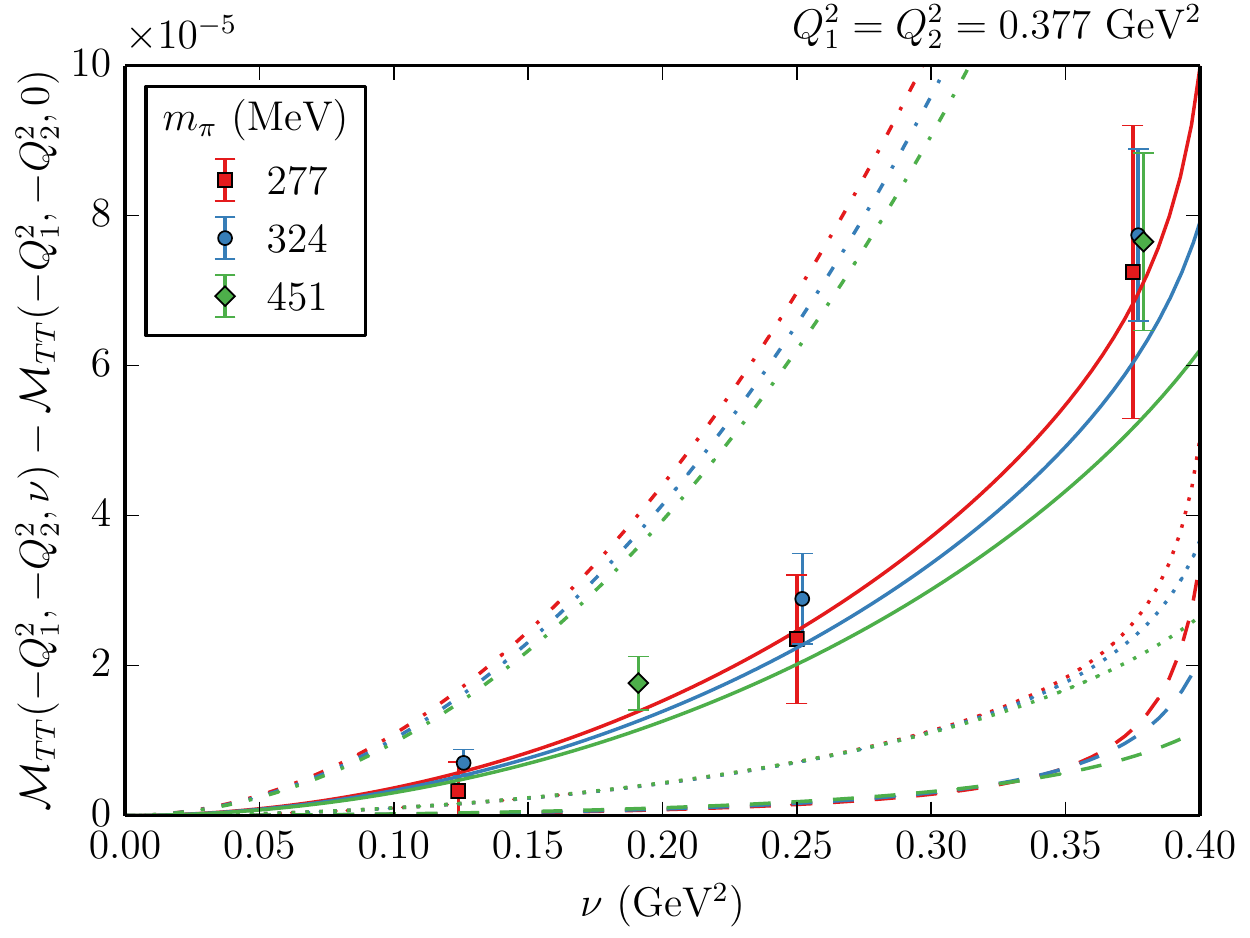}
  \caption{Forward scattering amplitude $\mathcal{M}_{TT}$ versus
    $\nu$, for fixed $Q_1^2$ and $Q_2^2$, and three different pion
    masses. The points are lattice data and the solid curves result
    from applying the dispersion relation
    [Eq.~(\protect\ref{eq:dispersion})] to the model for the
    $\gamma^*\gamma^*\to\text{hadrons}$ cross section. The dashed and
    dotted curves show the contributions from $\pi^0$ and
    $\pi^0+\eta'$, and the dash-dotted curves include an additional
    high-energy Regge contribution to the cross section for the case
    of real photons at the physical pion mass.}
  \label{fig:amp_fix_Q1_Q2}
  \end{minipage}
\end{figure}

We show a comparison of lattice data and the result from applying the
dispersion relation to our model in Figs.~\ref{fig:amp_fix_Q1_mpi} and
\ref{fig:amp_fix_Q1_Q2}. Figure~\ref{fig:amp_fix_Q1_mpi} shows the
wide range of $(Q_2^2,\nu)$ accessible in a single calculation at
fixed $Q_1^2=0.377\text{ GeV}^2$ and pion mass 324~MeV. The model
produces good agreement with the data, although it has a considerable
uncertainty. In particular, there is no data from experiment
constraining the two-photon transition form factors for scalar and
tensor mesons. We take this to be of the form
$F(Q_1^2,Q_2^2)=\prod_{i=1,2}1/(1+Q_i^2/\Lambda^2)$ and set $\Lambda$
by hand to 1.6~GeV to produce the nice agreement with our data;
however, varying $\Lambda$ by $\pm 0.4$~GeV causes the curves to
change by up to $\pm50\%$.

In Fig.~\ref{fig:amp_fix_Q1_Q2}, we fix both photon virtualities to
$0.377\text{ GeV}^2$ and show the dependence of $\mathcal{M}_{TT}$ on
$\nu$, for three different pion masses. We find no significant
dependence on the pion mass. The figure also shows the contributions
from $\pi^0$ and $\eta'$ final states in the model, which are far from
being the dominant contributions. In order to see the dominance of the
$\pi^0$ contribution, which is expected for $a_\mu^\text{HLbL}$, we
may need to reach much smaller photon virtualities and
closer-to-physical pion masses. Finally, we show in the figure a
high-energy contribution to the $\gamma\gamma\to\text{hadrons}$ cross
section, which is obtained using a fit to data based on Regge
theory. This is not included in the main model due to the lack of an
extrapolation to heavier-than-physical pion masses and nonzero photon
virtualities, but is an indication of possible additional uncertainty
in the model.

\section{Strategy for evaluation of the HLbL contribution to the muon $\mathbf{g-2}$}

In Euclidean space, we put the muon on shell with momentum
$p=im\hat\epsilon$ in an arbitrary direction $\hat\epsilon$
($\hat\epsilon^2=1$). Applying infinite-volume continuum QED Feynman
rules to evaluate the diagram in Fig.~\ref{fig:HVP_HLbL} (right) and
then isolating the electromagnetic form factor $F_2(0)$ for the muon,
we obtain an expression of the form
\begin{equation}
  a_\mu^\text{HLbL} = \int d^4x\, d^4y\,
  \mathcal{L}_{[\rho,\sigma];\mu\nu\lambda}(\hat\epsilon,x,y)
  i\hat\Pi_{\rho;\mu\nu\lambda\sigma}(x,y),
\end{equation}
where $\mathcal{L}$ is a kernel involving muon and photon propagators, and
\begin{equation}
  \hat\Pi_{\rho;\mu\nu\lambda\sigma}(x_1,x_2) =
  \int d^4x_4\, (ix_4)_\rho \langle
  J_\mu(x_1) J_\nu(x_2) J_\lambda(0) J_\sigma(x_4) \rangle
\end{equation}
is the four-point function with a momentum derivative applied to the
vertex that couples to the external photon. After contracting
$\mathcal{L}$ with $\hat\Pi$, the integrand for $a_\mu^\text{HLbL}$ is
a scalar function and thus depends only on the five invariants $x^2$,
$y^2$, $x\cdot y$, $x\cdot\hat\epsilon$, and $y\cdot\hat\epsilon$;
therefore, three of the eight dimensions in the integral are trivial.
The dimensionality can be further reduced by noting that the result is
independent of $\hat\epsilon$, which we can therefore eliminate by
averaging in the integrand. If we replace
\newcommand{\bL}{\overline{\mathcal{L}}}
\begin{equation}
  \mathcal{L}(\hat\epsilon,x,y) \to
  \bL(x,y) \equiv
  \langle\mathcal{L}(\hat\epsilon,x,y)\rangle_{\hat\epsilon},
\end{equation}
then the integrand will only depend on $x^2$, $y^2$, and $x\cdot y$.

Putting the ingredients together, we arrive at our formula:
\begin{equation}
  a_\mu^\text{HLbL} = 2\pi^2 \int_0^\infty |x|^3d|x| \int d^4y\,d^4z\,
  \bL_{[\rho,\sigma];\mu\nu\lambda}(x,y)(-z)_\rho
  \langle J_\mu(x) J_\nu(y) J_\lambda(0) J_\sigma(z) \rangle,
\end{equation}
where we have used the fact that, after integrating over $y$ and $z$,
the integrand depends only on $x^2$ so that the integral over $x$ can
be reduced to one dimension. We plan to compute the integrals over $y$
and $z$ similarly to our computation of the momentum-space four-point
function. First, we will use local currents at $0$ and $x$, and
compute point-source propagators from those points. Next, we will
implement the integral over $z$ using sequential propagators with
conserved-current insertions, and obtain $\hat\Pi(x,y)$ at all points
$y$. We will then contract it with $\bL(x,y)$ to integrate over
$y$. These steps will have a similar cost to our previous evaluation
of the scattering amplitude with two of the momenta fixed. Finally, we
will repeat the process several times to perform the one-dimensional
integral over $|x|$.

\section{Summary and outlook}

We have shown that the contribution from the fully quark-connected
four-point function to the HLbL scattering amplitude can be
efficiently evaluated on the lattice if two of the three momenta are
fixed. In the case of forward scattering, this amplitude is related
model-independently by a dispersion relation to the
$\gamma^*\gamma^*\to\text{hadrons}$ cross section. Using a model for
the latter, we found that our lattice results are consistent with
phenomenology, within the large model uncertainties. For typical
Euclidean kinematics, we do not find a dominant contribution from
neutral pion exchange.

Our strategy for computing the leading-order HLbL contribution to the
muon $g-2$ is in place; work is ongoing to evaluate the kernel
$\bL_{[\rho,\sigma];\mu\nu\lambda}(x,y)$. Since several model
calculations~\cite{Jegerlehner:2009ry,Pauk:2014rta,Bijnens:2015jqa}
indicate that neutral pion exchange is the leading contribution to
$a_\mu^\text{HLbL}$, it may be challenging to reach the physical
regime on the lattice --- near-physical pion masses and large volumes
may be called for. In addition, disconnected diagrams may be
important, as Ref.~\cite{Bijnens:2015jqa} argues that omitting them
will lead to an unphysical enhanced $\pi^0$ contribution appearing
instead of the $\eta'$ contribution.

Finally, we also note the recent computation of the connected
contribution to $a_\mu^\text{HLbL}$ using non-perturbative lattice QCD
and perturbative lattice QED~\cite{Blum:2015gfa}.

\acknowledgments

We thank our colleagues within CLS for sharing the lattice ensembles
used. This work made use of the ``Clover'' cluster at the Helmholtz
Institute Mainz; we thank Dalibor Djukanovic for technical
support. The programs were written using QDP++~\cite{Edwards:2004sx}
with the deflated SAP+GCR solver from openQCD~\cite{OpenQCD}. J.G.\
thanks the Institute for Nuclear Theory at the University of
Washington for its hospitality and the Department of Energy for
partial support during the writing of this document.

\bibliographystyle{JHEP-2}
\bibliography{hlbl}

\end{document}